\documentclass{WileyMSP-template}

\usepackage{booktabs}
\usepackage{amsmath}
\usepackage[table]{xcolor}

\newcommand{\srf}{\ensuremath{\mathrm{SrF}_2}}

\begin{document}


\title{\noindent Control of the Electronic and Atomic Structure of the Altermagnetic $\alpha$-MnTe Surface  via Substrate-Induced Surfactants

}

\maketitle


\author{Martin Heinrich,$^{1,2}$*}
\author{Mahdi Hajlaoui,$^{2}$}
\author{Takuya Takashiro,$^{3}$}
\author{Tetiana Zakusylo,$^{2}$}
\author{Petr Babor,$^{4}$}
\author{Ondřej Caha,$^{5}$}
\author{Dawid Wutke,$^{6}$}
\author{Natalia Olszowska,$^{6}$}
\author{Gunther Springholz,$^{2}$*}
\author{Matthias Muntwiler$^{1}$*}



\begin{affiliations}
\vspace{1mm}

\noindent $^{1}$ Swiss Nanoscience Institute, University of Basel, 4056 Basel and
Paul Scherrer Institute, Center for Photon Science, 5232 Villigen PSI, Switzerland \\ 
Email Address: matthias.muntwiler@psi.ch\\
\noindent $^{2}$ 
Johannes Kepler University, Institute of Semiconductor and Solid State Physics, 4040 Linz, Austria, 
Email Address: martin.heinrich@jku.at, and gunther.springholz@jku.at \\
\noindent $^{3}$ 
Tohoku University, Department of Physics, Graduate School of Science, Sendai 980-8578, Japan\\
\noindent $^{4}$ 
Brno University of Technology, Department of Surfaces and Nanostructures, 60200 Brno, Czech Republic\\
\noindent $^{5}$ 
Masaryk University, Department of Condensed Matter Physics, 60177 Brno, Czech Republic\\
\noindent $^{6}$ 
National Synchrotron Radiation Centre SOLARIS, Jagiellonian University, 30-392 Kraków, Poland

\end{affiliations}



\begin{abstract}
Altermagnets are a novel class of materials with unique momentum dependent spin-polarization of the electronic band structure, which opens up new avenues for spintronic devices. Altermagnetic $\alpha$-MnTe is particularly interesting for such applications due to its giant spin splitting and the persistence of altermagnetic order up to room temperature. Growth of thin films is, however, challenging and governed by interactions with the substrate. As a result, the surface of $\alpha$-MnTe thin films exhibits very different atomic structures in dependence of the substrate material. 
Whereas on SrF$_2$(111) an atomically clean, perfectly (2$\times${}2) reconstructed surface forms, on InP(111)A the MnTe surface is always InTe terminated due to super-segregation of In from the substrate to the surface. Most strikingly, the InTe surfactant layer displays a quasi-periodic dislocation network to accommodate the InTe/MnTe lattice mismatch. This leads to the emergence of distinct surface states in dependence of surface termination that strongly affect the electronic band structure at the surface as shown by angle resolved photoemission and tunneling spectroscopy. Thus, control and characterization of the atomic structure of the surface is fundamentally important for understanding the properties of altermagnets and provides room for engineering and tailoring of spin-polarized states in actual devices.
\end{abstract}

%
%
%
%

\section{Introduction}
The recent discovery of altermagnetism as a distinct new magnetic phase by Šmejkal et al. \cite{SmejkalL22PRX} has triggered tremendous theoretical and experimental efforts to explore its physical manifestations and to identify promising material candidates.
The concept is based on symmetry considerations well established in topology and superconductivity, and applies them to  unconventional forms of magnetism that combine a unique spin-polarized band structure with zero net magnetization. Consequently, altermagnets open up promising routes for power efficient electrical control of spin currents in spintronic devices via charge-to-spin conversion, anomalous Hall effect, and giant magnetoresistance  \cite{SmejkalL22PRXb,SongC25NRM}.

\begin{figure}[t]
	\centering
	\includegraphics[width=1.0\linewidth]{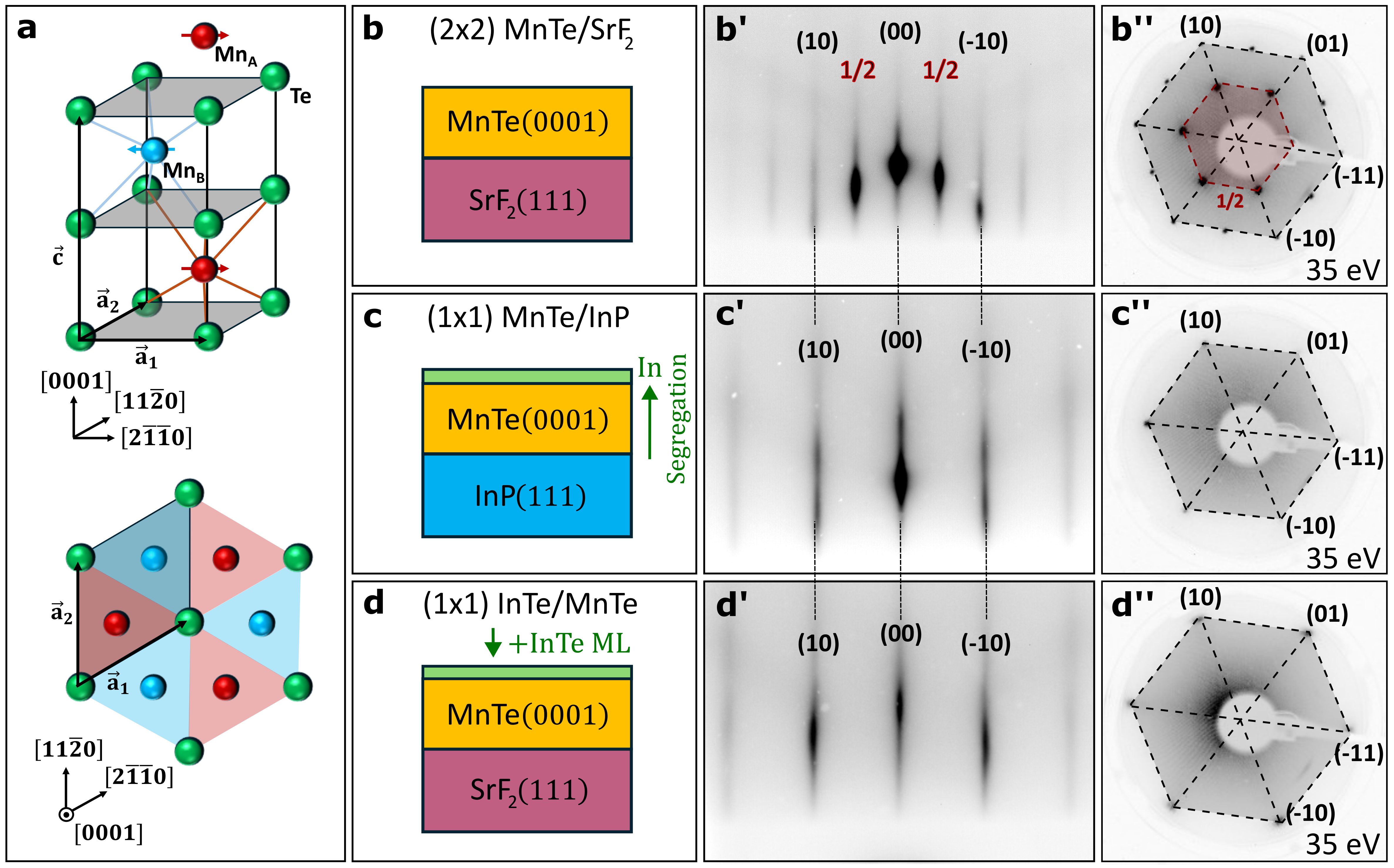}
	\caption{(a) Bulk crystal structure and altermagnetic spin texture of hexagonal $\alpha$-MnTe (top) together with its projection on the (0001) surface (bottom). The bulk lattice parameters are $a$ = $4.14$ \AA\ and $c$ = $6.7$ \AA\ at room temperature \cite{KriegnerD17PRB}. The investigated types of layer structures are shown in panels (b-d), consisting of MnTe on \srf{}(111), MnTe on InP(111)A, and MnTe on \srf{}(111) with an additional atomic layer of InTe  deposited on top. This consistently results in different diffraction patterns as evidenced by in situ RHEED (b'-d') \emph{during} growth, as well as by ex situ LEED (b''-d'') measured after cool down to room temperature. For MnTe on \srf{}(111) 1/2 order diffraction spots appear indicating the formation of a (2$\times${}2) surface reconstruction. These are missing for the other cases, where the surface remains in a (1$\times${}1) state. RHEED was recorded along the $[11\overline{2}0]$ azimuth direction, corresponding to $[\overline{1}10]$ for the cubic substrates, LEED at 35 eV electron beam energy.
	}
	\label{fig:intro}
\end{figure}

Among the material candidates, altermagnetic $\alpha$-MnTe stands out due to its exceptionally large spin splitting \cite{SmejkalL22PRXb, KrempaskyJ24N,Hajlaoui2024} and robust magnetic ordering up to room temperature \cite{KriegnerD16NC,KriegnerD17PRB}.
It crystallizes in the hexagonal NiAs-structure with P$6_3$/mmc space group symmetry, where the magnetic moments of the Mn atoms align parallel within the basal c-planes but stack antiparallel along the c-direction (see  \textbf{Figure \ref{fig:intro}a}) \cite{KriegnerD16NC,KriegnerD17PRB,MazinI23PRB}. 
Although in this structure the magnetic moments are fully compensated, the spin sublattices connect via a non-symmorphic six-fold screw-axis rotation $C_{6z}t_{1/2}$ and $C_{2y}$ spin rotation. 
This time-reversal symmetry breaking leads to a  momentum-dependent spin splitting and spin polarization of the electronic structure even without considering relativistic spin-orbit coupling \cite{SmejkalL22PRX, SmejkalL22PRXb,KrempaskyJ24N, betancourt_mnte}. 

To identify its altermagnetic characteristics, various studies on the anisotropic magnetoresistance (AMR) \cite{KriegnerD16NC}, anomalous Hall effect \cite{KriegnerD17PRB,AminO24N,HarikiA24PRL} and 
circular and linear dichroism in X-ray absorption spectroscopy (XAS) \cite{HarikiA24PRL,LeeS24PRL} of $\alpha$-MnTe have been performed. 
Moreover, the spin-split band structure and broken Kramers degeneracy were confirmed by angle-resolved photoemission spectroscopy (ARPES) \cite{KrempaskyJ24N,LeeS24PRL,OsumiT24PRB}, though single-domain spin-resolved measurements remain challenging due to the intrinsic multidomain structure \cite{KriegnerD16NC,KriegnerD17PRB}, as has been revealed by photoemission electron microscopy (PEEM) utilizing X-ray magnetic linear and circular dichroism (XMLD, XMCD) \cite{AminO24N}.

Importantly, while altermagnetism is induced by the non-symmorphic symmetry of the bulk, this symmetry is inherently broken at the surface. This has triggered an intense debate on how the termination and atomic arrangements at the surface alter the properties of altermagnets.
This is particularly important for thin film devices, whose properties may be completely dominated by surface effects.  
Indeed, recent studies highlight the profound impact of the surface on both the altermagnetic 
spin texture \cite{zeng2025nonaltermagneticspintexturemnte} and the emergent anomalous Hall effect 
\cite{zhao2026emergentanomaloushalleffect, zhou2026surfacestatedrivenanomaloushalleffect}.
Thus, understanding and controlling the surface properties of altermagnets is crucial for device applications.

\
\vspace{1 mm}

In this work work, we demonstrate that the electronic structure of altermagnetic $\alpha$-MnTe(0001) 
strongly depends on the real space atomic structure of the surface that differs significantly  from the bulk. 
Using a multiscale approach combining molecular beam epitaxy (MBE), scanning tunneling microscopy (STM), low-energy electron diffraction (LEED) and x-ray photoelectron spectroscopy (XPS), we reveal that this arises from strong substrate/film interactions and extreme surface segregation of substrate atoms. 
For MnTe films on InP substrates, it induces the formation of a self-assembling InTe surfactant layer that continuously rides on top of the growing film, irrespective of thickness.
Most strikingly, this surfactant layer forms a highly regular, hexagonal misfit dislocation network for strain relief, imposing a large nanoscale modulation of strain and local electronic structure at the surface.

Electronic structure measurements using ARPES unveil the emergence of distinct surface states superimposed on the bulk bands. This even applies to the atomically clean MnTe(0001) surface on \srf(111) with its intrinsic (2$\times${}2) surface reconstruction that halves the surface Brillouin zone (SBZ) relative to the bulk projection.
This leads to the general conclusion that surface reconfigurations add significant weight to the local density of states (LDOS) of altermagnetic MnTe. This alters the spin textures and spin-polarization probed by surface-sensitive techniques like ARPES and directly impacts the transport characteristics of thin-film devices. Finally, while this study focuses on MnTe on InP and \srf{}, we emphasize that similar effects are expected for films grown on other substrates such as GaAs or van der Waals bonded V-VI chalcogenide buffer layers, offering a new paradigm for engineering altermagnetic heterostructures.

\vspace{3 mm}

\section{Results and Discussion}

\subsection{Growth and surface properties}

\begin{figure}[t]
	\centering
	\includegraphics[width=0.84\linewidth]{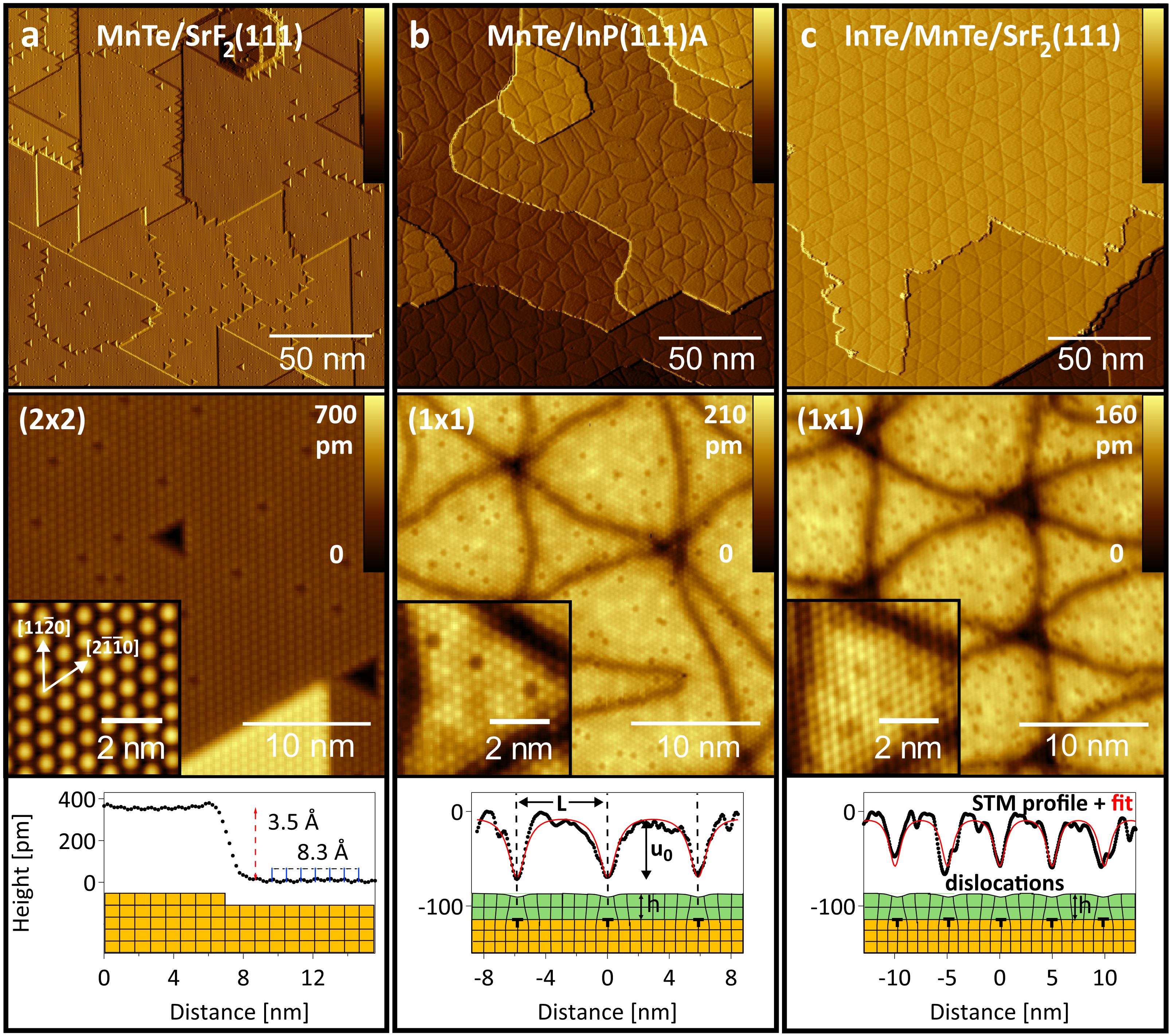}
	\caption{\textbf{STM surface images} of the different types of MnTe layers shown in Figure 1. (a)  MnTe/\srf{}(111) with (2$\times${}2) surface reconstruction, (b) MnTe/InP(111)A with (1$\times${}1) reconstruction and (c) MnTe/\srf{}(111) with one monolayer of InTe deposited on top. The top row displays derivative overview images demonstrating perfect 2D layer-by-layer growth in each case. The different atomic arrangements at the surfaces are revealed by the higher magnification images displayed in the middle row. Height profiles measured along the $[11\overline{2}0]$ direction are shown at the bottom together with the schematic illustration of the surface structures that indicate the origin of the dislocation patterns in (b) and (c). STM measurements were performed at $T = 4.4\,\mathrm{K}$, $V_{\mathrm{tip}} = 1-1.5\,\mathrm{V}$, $I_t = 100-200 \,\mathrm{pA}$.
    }
	\label{fig:stm}
\end{figure}

MnTe has been grown as crystallites using self-flux \cite{KrempaskyJ24N} or vapor transport \cite{OsumiT24PRB} methods or as thin films by molecular beam epitaxy (MBE) \cite{KrempaskyJ24N,KriegnerD16NC,LeeS24PRL,SpringholzG13,SpringholzG14}.
Epitaxy yields high-quality thin films with large crystalline domains, precise composition and a well-defined surface structure, but requires lattice-matched substrates to minimize strain that eventually leads to distortions and defects.
For hexagonal $\alpha$-MnTe(0001), InP(111)A as well as \srf{}(111) substrates fulfill this condition because their surface lattice parameter of $a = a_0/\sqrt{3}$ = 4.10 Å and 4.15 Å, respectively, are very close to the hexagonal in-plane lattice constant of $\alpha$-MnTe (0001) of 4.14 Å. 
As shown by our previous works \cite{KrempaskyJ24N,KriegnerD16NC,KriegnerD17PRB}, high quality epilayers are obtained in both cases, exhibiting an onset of magnetic ordering at $T_N \approx$ 306 K \cite{KriegnerD16NC,KriegnerD17PRB} and pronounced altermagnetic behavior in magnetotransport experiments \cite{KriegnerD16NC,betancourt_mnte}.

However, the surface and bonding properties of \srf{} and InP(111) are very different. On the one hand, \srf{} is a highly ionic insulator but its very low-energy natural (111) cleavage surface is rather inert. This results in a weak film/substrate interaction, for which reason MnTe initially starts in a Vollmer-Weber growth such that a flat 2D surface is regained only after about 100 -- 200 nm MnTe deposition.
InP, on the other hand, is a covalently bonded III-V semiconductor that, due to its dangling bonds, interacts much stronger with MnTe. Thus, a 2D wetting layer is immediately formed upon MnTe deposition and the growth proceeds in a purely Frank van der Merwe mode. Moreover, it is noted that the polar (111)A surface is indium terminated, which is essential for obtaining MnTe in the hexagonal phase. 

For our present investigation, MnTe films were grown on both substrates (Figure~\ref{fig:intro}b-c) using \textit{identical} conditions, i.e., a substrate temperature of 360$\degree$C and  Te:Mn beam flux ratio of 3 as detailed in the experimental section. The film thickness was 200 -- 400 nm in both cases.
To study the effect of indium segregation, additional growth experiments were performed in which 1 -- 2 monolayers (ML) of indium were deliberately deposited on top of the surface as illustrated by Figure~\ref{fig:intro}d.

To our great surprise, although the lattice structure of the MnTe films on \srf{}(111) and  InP(111)A as seen in X-ray diffraction is identical \cite{KriegnerD16NC,KriegnerD17PRB}, a completely different surface structure is observed. This is revealed by the distinct RHEED and LEED diffraction patterns displayed on the right hand side of Figure~\ref{fig:intro}, where consistently half-order diffraction spots are observed for the MnTe/\srf{}(111) surface, whereas for the MnTe/InP(111)A surface only integer Bragg peaks appear. Thus, MnTe on \srf{}(111) forms a (2$\times${}2) surface reconstruction, whereas on InP the surface always remains unreconstructed in a (1$\times${}1) phase, independent of growth condition and layer thickness. Most notably, a (1$\times${}1) structure also emerges when one monolayer (ML) of InTe is deposited on the (2$\times${}2) MnTe on \srf{} surface (see Figure~\ref{fig:intro}d' and d'' and the supplemental information (SI)).

\subsection{Atomic structure and composition of the MnTe surfaces}

\begin{figure}[t]
	\centering
	\includegraphics[width=0.96\linewidth]{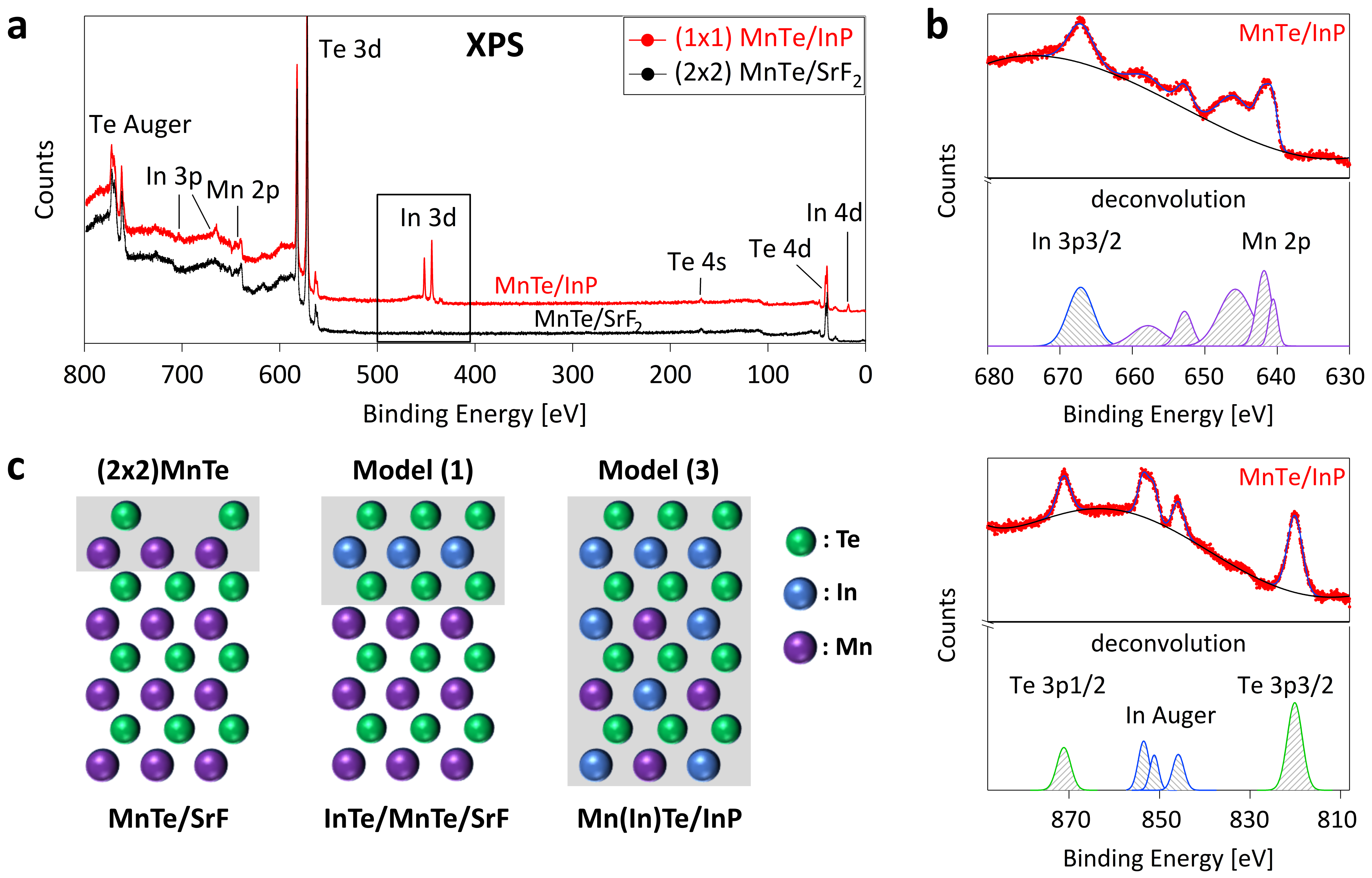}
	\caption{\textbf{Surface composition derived by XPS:} (a) Overview spectra of the MnTe/InP(111)A (red) and MnTe/\srf{}(111) (black) film surfaces, and (b) detailed core level spectra of the Mn 2p, Te 3p and In 3p peaks of the MnTe/InP(111)A sample. The overview spectrum of MnTe/InP(111)A has been purposely up-shifted for better comparison. (c) Surface layer models for the  (2$\times${}2) MnTe, InTe/MnTe and InTe/$\mathrm{In}_x \mathrm{Mn}_{1-x} \mathrm{Te}$ surfaces described in the text. Grayed areas indicate key differences between the different models.}
	\label{fig:xps}
\end{figure}

To determine the real space atomic structure of the surfaces, STM imaging was performed for all samples as shown in \textbf{Figure~\ref{fig:stm}}. Evidently, in all cases, the film surfaces are atomically flat and consist only of wide terraces confined by single monoatomic steps. 
For MnTe on \srf{}(111) shown in (a), the terrace edges are faceted with straight step edges aligned along the six-fold $\langle 11\overline{2}0 \rangle$ directions. The height of these steps is 3.5 Å, which corresponds to the thickness of one monolayer that is equal to half of the vertical MnTe lattice constant $c$. On the atomic scale, however, the hexagonal atom lattice exhibits a periodicity of 8.3 Å, which is  twice as large as the in-plane lattice constant $a$ = 4.14 Å of bulk MnTe \cite{GreenwaldS53AC,SzuszkiewiczW06PRB} (high resolution STM images in the lower part of Figure~\ref{fig:stm}a). This means that on the surface only every second lattice site is occupied, which perfectly agrees with the (2$\times${}2) surface reconstruction observed by LEED and RHEED (Figure~\ref{fig:intro}b' and b''). Since STM was performed  with positive tip bias (filled state imaging), the visible atoms in the STM images are Te adatoms because the valence band is predominantly formed by the Te anion states \cite{mnte_bands-orbitals}.
This Te-termination is in line with our observation that prolonged post-growth annealing of the (2$\times${}2) surface leads to a desorption of Te by which it eventually evolves into a defect-rich (1$\times${}1) Mn-terminated surface, which is not discussed here.
Since our epitaxial layers grow under excess Te flux and Te at the surface tends to be energetically more favorable for most telluride compounds, we conclude that the (2$\times${}2) surface is terminated by tellurium, as sketched in Figure~\ref{fig:xps}c.

For MnTe on InP(111)A, the surface structure looks completely different. As revealed by the STM images of Figure~\ref{fig:stm}b not only the terrace edges are no longer straight, but most strikingly, a dense network of dark lines appears on the surface. Their average separation of 10 -- 14 nm is much larger than the spacing of the surface atoms, and -- as shown by the surface profiles depicted in the lower panel of Figure~\ref{fig:stm}b -- they correspond to local vertical surface depressions by about 0.6 \AA{}. On the atomic scale a continuous hexagonal lattice is preserved, albeit with a much higher number of atomic defects than for the MnTe/\srf{} case (see insets of Fig.~\ref{fig:stm}). Interestingly, the triangular network of dark lines is strongly reminiscent to those observed in highly mismatched heteroepitaxial thin film systems such as Ge/Si(111) \cite{FILIMONOV200576}, InAs/GaAs(111) \cite{nosho_InAs-GaAs} or InN/GaN(0001) \cite{Liu_disloc}, where these dark lines arise due to misfit dislocations at the heterointerfaces. This suggests that on top of the MnTe/InP surface there resides a thin overlayer with a large lattice mismatch to MnTe that is accommodated by subsurface dislocation formation. The observed dark lines are therefore due to the local distortions of the crystal lattice around  each dislocation (see Refs. \cite{SPRINGHOLZ199712,wiesauer_disloc}) as illustrated schematically in the bottom panel of Figure~\ref{fig:stm}b.
Because the average periodicity of the dislocation network is roughly 30 times larger than the in-plane lattice parameter of MnTe, the lattice constant of the overlayer must be at about 3\% larger than that of MnTe (see Section 2.3). This agrees with the fact that the surface lattice constant determined from the atom-resolved STM image amounts to 4.2 Å, which is larger compared to bulk MnTe (4.14 \AA). Nevertheless, this atom spacing corresponds to a (1$\times${}1) structure, which explains why no half-order diffraction spots are present in LEED and RHEED as shown in Figure~\ref{fig:intro}b' and b''.  

To verify the hypothesis of the existence of a strained overlayer on the surface, we have performed a control experiment in which we deposited one monolayer of InTe onto the dislocation-free (2$\times${}2) MnTe/\srf{} surface. The resulting STM images in Figure~\ref{fig:stm}c demonstrate that \emph{exactly} the same dislocation network forms, essentially with the same triangular structure and line spacings. This striking similarity proves that for MnTe films grown on InP, the surface is intrinsically InTe terminated, despite the fact that for the given film thickness of 200 nm, the MnTe/InP interface is about 2000 atomic layers away from the surface. 

The marked difference between the surface termination of the MnTe/InP(111)A and MnTe/\srf{} films is confirmed by XPS measurements shown in \textbf{Figure~\ref{fig:xps}}a. For the MnTe/ InP(111)A films, In peaks appear alongside the expected Te and Mn peaks, whereas for the MnTe/\srf{} sample only Mn and Te peaks with an almost stoichiometric 1:1 ratio (cf.\ supplementary information (SI) Figure S3) appear. Consequently, the MnTe/\srf{} sample is atomically clean, whereas the quantitative analysis of the Mn 2p, Te 3p and In 3p core levels (Figure~\ref{fig:xps}b) shows that for MnTe/InP(111)A the In coverage of the surface is slightly above one atomic monolayer. This is verified by our control experiment on the 1 ML MnTe deposited on MnTe/\srf{}, where the same In core level peaks appear in the XPS spectra with similar intensity ratios (see SI for the details).  

Taking everything together, our findings clearly reveal that an ultra-thin but extremely stable InTe surfactant layer forms on MnTe films on InP(111)A substrates which is completely absent when the growth is performed under indium-free conditions. This extraordinary effect suggests that the InTe termination  drastically lowers the free surface energy of MnTe(0001). This scenario is strongly supported by the fact that the In-induced (1$\times${}1) surface reconstruction remains present even for micrometer thick films. In fact, this is also observed for 1 ML InTe on MnTe/\srf{} films, in which case the original (2$\times${}2) reconstruction does not recover even after depositing additional several hundred nanometers of MnTe (see supplemental information). The remarkable stability of the InTe surfactant is further evidenced by post-growth annealing experiments, showing that the clean (2$\times${}2) MnTe(0001) phase can only be recovered by high temperature annealing above 600 $\degree$C  where the InTe surfactant layer desorbs from the MnTe surface. This process, however, results in a significantly degraded surface (see supplemental information). A clean and well ordered MnTe surface can only by achieved on \srf{}(111).

\subsection{Surface deformations and dislocation network}

\label{sec:dislocation_network_model}

To better understand the peculiar structure of the InTe surfactant layer, we need to consider the line defects formed at the InTe/MnTe interface. In heteroepitaxial layers, such line defects are formed in order to relieve the lattice mismatch to the underlying substrate by interfacial misfit dislocations. Most efficient for strain relaxation are edge dislocations with the \emph{Burgers vector} $\vec{b}$ parallel to the interface. In general, this Burgers vector corresponds to a full atomic lattice shift, called \emph{perfect dislocations}, or to a sequence of shorter shifts, called \emph{partial dislocations}. 
Although each misfit dislocation relieves a certain amount of strain, it locally induces a distortion of the surrounding lattice planes due to the broken bonds at the dislocation core (see Figure~\ref{fig:stm}b \& c). As shown by our previous work \cite{SPRINGHOLZ199712}, this leads to a local surface deformation above each misfit dislocation whose amplitude and shape depends on the absolute value and direction of the Burgers vector.  

For a single edge dislocation located at a certain depth $h$ below the surface and $\vec{b}$ parallel to the interface, the resulting surface deformation, i.e., vertical displacement $u_z$ can be calculated by elasticity theory \cite{SPRINGHOLZ199712} as:
\begin{equation}
    u_{z}(x)=\frac{b_{edge}}{\pi}\frac{h^2}{x^2 + h^2},
    \label{eq:singledislo}
\end{equation}
where $x$ is the lateral distance from the dislocation line and $b_{edge}$ is the edge component of the Burgers vector. Thus, this local surface depression has an amplitude of $u_{z,0}={b_{edge}}/{\pi}$ and a lateral width that scales $\propto h$. Since $b$ typically corresponds to one unit cell vector, the amplitude of the depression is of the order of 1 \AA , which agrees to what we observe by STM (Figure~\ref{fig:stm}). For a periodic edge dislocation network, this solution can be extended to an infinite array of dislocations as \cite{FILIMONOV200576, andrews_disloc}:
\begin{equation}
    u_{z}(x)=\frac{b\overline{h}\sinh{(2\pi\overline{h})}}{\cosh{(2\pi\overline{h})}-\cos{(2\pi\overline{x})}},
    \label{eq:multidislo}
\end{equation}
where $\overline{h}=\frac{h}{L}$ and $\overline{x}=\frac{x}{L}$ are the dislocation depth (film thickness) and the lateral distance normalized to the mean dislocation spacing $L$, respectively. 
With this model, the STM profiles across the InTe terminated MnTe surfaces can be well reproduced as shown by the lowest panels of Figure~\ref{fig:stm}b and c, with an excellent agreement between theory (red lines) and experiments (black dotted lines), using the measured dislocation spacing as input parameter $L$.    

For the measured amplitude of the surface depressions of $u_0=60 \pm 4$ pm, the model yields the edge component of the Burgers vector as $b_{edge} = 1.9\pm 0.2$ \AA, which suggests that the triangular dislocation network consists of $90\degree$ partial dislocations with $\lvert \vec{b} \lvert = a /\sqrt{6}$. These are known to be energetically more favorable than perfect dislocations in hexagonal and cubic (111) oriented  heteroepitaxial systems \cite{Xu_disloc}. 
Moreover, the full fit of the deformation shape indicates that the dislocations are at about 6.8 \AA{} below the surface. This suggests that the thickness of the InTe surfactant layer is roughly equal to one hexagonal unit cell (Figure~\ref{fig:intro}a). This should be regarded as an estimate because in the model (Equations~\ref{eq:singledislo} and \ref{eq:multidislo}), elastic anisotropy as well as nonlinear corrections needed to describe the displacement in the vicinity of the dislocation cores are neglected. 

Since the dense dislocation network obviously accommodates a large amount of the InTe/MnTe lattice mismatch, we can estimate this mismatch by the relation $\Delta a / a = 1.5 \cdot b_{edge}/L_{\parallel}$ \cite{wiesauer_disloc}. Using $L_{\parallel} = 12\pm2$ nm determined by STM, the lattice parameter of the InTe surfactant must be  at least 2\% larger than that of MnTe. 
Bulk InTe, however,  crystallizes in a tetragonal crystal structure that is quite incompatible with the hexagonal MnTe lattice, but under hydrostatic pressure a rock salt InTe phase exists with a lattice constant of 6.15 \AA, corresponding to  $a \approx 4.35\,\mathrm{\AA}$ at the surface \cite{zhou2026surfacestatedrivenanomaloushalleffect,doi:10.1126/science.142.3593.662}, reasonably close to $a$ of $\alpha$-MnTe. 
Moreover, for rock salt structures the atomic layer stacking along [111] is well compatible with the stacking of the hexagonal MnTe lattice. 
This way, it is plausible that the rock salt structure of InTe is stabilized by the lattice matching to the MnTe film, where part of the strain is relieved by the dislocations. 
It is further noted that a stable van der Waals type tetralayer structure has been proposed for free-standing InTe layers \cite{zolyomi2014, giovannetti_kekule, Ayadi2017, Lee2024}. 
However, the corresponding structure model (5) in Table~\ref{tab:xps-data-models} is not compatible with the XPS data, indicating that the binding to MnTe is much stronger than van-der-Waals forces.
More details are given in the supplemental material.   

\subsection{Surface  modeling and indium distribution}

\begin{table}[t]
    \centering
	\caption{{\textbf{Surface elemental compositions}} extracted from XPS (Figure~\ref{fig:xps}a and SI Figure S3) measured under low and high emission angles $\theta$. The details of the analysis are described in the supplemental information. This is compared to simulations for different atomic surface models as exemplified in Fig. 3(c). Colored rows indicate well-matching models and samples.}
	\begin{tabular}{clccccc}
		\toprule
		\multicolumn{2}{l} { I. Surface composition obtained by XPS } & $\theta$ & \%Te & \%Mn & \%In & In/Mn  \\
		\midrule
        (A) &  MnTe/\srf{} with (2$\times${}2) reconstruction & $20\degree$ & $48.5$ & $51.5$ & 0 & 0 \\
		& & $60\degree$ & 47.9 & 52.1 & 0 & 0 \\
		\rowcolor{lightgray} 
        \midrule
        (B) & MnTe/InP with (1$\times${}1) segregated InTe surface layer & 20\degree & 43.9 & 35.4 & 20.7 & 0.58 \\
		\rowcolor{lightgray} 
        & & $60\degree$ & 45.7 & 28.7 & 25.6 & 0.89 \\
		\rowcolor{cyan} 
        \midrule
        (C) &  MnTe/\srf{} with (1$\times${}1) one monolayer InTe deposited on top & $20\degree$ & 49.2 & 36.7 & 14.1 & 0.38 \\
		\rowcolor{cyan} 
        & & $60\degree$ & 51.7 & 31.2 & 17.1 & 0.55 \\
		\midrule
        \multicolumn{2}{l}{ II. Model calculations for different atomic layer stackings (Fig. 3c)}  &  &  &  &  &  \\
        \midrule
        \rowcolor{cyan} (1) & Te-In-Te-/ MnTe \ldots & $0\degree$ & 53.6 & 33.6 & 12.8 & 0.38 \\
        \rowcolor{cyan} & & $60\degree$ & 56.2 & 26.2 & 17.6 & 0.67 \\
        (2) & Te-In-Mn-Te-/ MnTe \ldots & $0\degree$ & 48.4 & 38.8 & 12.8 & 0.33 \\
        & & $60\degree$ & 48.8 & 33.6 & 17.6 & 0.52 \\
        \rowcolor{lightgray} (3) & Te-In-Te-/ $\text{Mn}_{0.88}\text{In}_{0.12}$ \ldots & $0\degree$ & 53.6 & 29.5 & 16.9 & 0.57 \\
        \rowcolor{lightgray} & & $60\degree$ & 56.2 & 23.1 & 20.7 & 0.90 \\
        (4) & Te-In-$\text{Mn}_{0.88}\text{In}_{0.12}${}-Te \ldots & $0\degree$ & 48.4 & 34.1 & 17.5 & 0.51 \\
        & & $60\degree$ & 48.8 & 29.6 & 21.6 & 0.73 \\
        (5) & Te-In-In-Te-Mn-Te-/ MnTe \ldots & $0\degree$ & 48.4 & 27.6 & 24.0 & 0.87 \\
        & & $60\degree$ & 48.8 & 20.0 & 31.2 & 1.56 \\
        \bottomrule
        
	\end{tabular}
	\label{tab:xps-data-models}
\end{table}

To determine whether indium completely segregates to the surface or dissolves homogeneously in the MnTe crystal, we measure XPS at two different emission angles: Near normal emission ($\theta = 20\degree$ to the surface normal) and near grazing emission ($\theta = 60\degree$) which is more surface sensitive.
Raw data and the processing steps are detailed in the supplemental material.
The relative weights of the XPS peaks, normalized by cross-section, are listed in the upper part of Table~\ref{tab:xps-data-models}.
First, we note that the MnTe/\srf{} sample shows approximately 50\% normalized intensity for each of the nominal  Mn and Te components, independent of angle and probing depth.
For the `MnTe/InP' sample we observe a large In concentration exceeding half of the Mn concentration.
Its relative intensity increases for grazing emission while Mn decreases, indicating a localization of In at the surface. Conversely, Te is distributed homogeneously, as its relative amount only varies by less than 2\% with probing depth.
Since the Te content is close to 50\% for both measurement geometries, we infer that the composition is $\mathrm{In}_{x}\mathrm{Mn}_{1-x}\mathrm{Te}$ throughout the sample and that only $x$ varies between 0 and 1 as a function of depth. 
In other words, the surface layer does not consist of pure In but of InTe, which is consistent with the fact that the growth is performed under Te-rich conditions.
Note that for MnTe/\srf{} sample with 1 ML InTe deposited on top, we observe similar quantities, though the In concentration and In/Mn ratio are a bit lower than for the MnTe/InP case, indicating a slightly larger thickness of the InTe surfactant layer on the latter surface. 

For a quantitative assessment, we devise different atomic models consisting of hexagonal In and Te layers stacked on top of the (0001) crystal truncation of $\alpha$-MnTe as illustrated in Figure~\ref{fig:xps}c.
To model the signal strength of the XPS peaks, we consider the probing depth according to the mean free path of the photoelectrons, which for the two measurement configurations is roughly two vertical lattice constants $c$ (8 atomic layers or 4 MnTe layers) at $\theta = 20\degree$ and slightly more than one lattice constant (5 atomic layers or 2.5 MnTe layers) at $\theta = 60\degree$ according to the universal curve \cite{Seah79}. Accordingly, the models are set to consist of 16 atomic layers in total.
The contribution of sub-surface layers is captured by an exponential weighting factor such that the contribution from the $16^{\text{th}}$ layer is reduced to $\frac{1}{e^2}$ at normal emission, and the atom densities within each layer are set equal to bulk MnTe.
Simple atom counting and exponential weighting then yields the model percentages that are compared to the experiments.

Out of over 20 models, the closest matches to the XPS data are listed in the lower part of Table~\ref{tab:xps-data-models}.
For the 1 ML InTe on MnTe sample (sample (C)), we find that the measured concentration of In closely matches the calculation for the single monolayer model (1). The MnTe film grown on InP (sample (B)) contains more In in the surface layers than predicted by model (1). This higher In/Mn ratio can be explained by a small incorporation of In also in the sub-surface Mn layers with $12\%$ of In as assumed by model (3). However, the XPS modeling does not yield a clear information about the deeper layers due to the limited probing depth. 

To determine if In incorporates also in the bulk of the films, we employed secondary ion mass spectroscopy measurements (SIMS) for depth profiling of composition. As shown in the supplemental material, the  measurements confirm the strong accumulation of In at the surface, but reveal that below the surface the In concentration rapidly decreases within a few nm to below the 1\% level, such that within the bulk of the MnTe film the In concentration is negligible. This indicates that equilibrium solubility of In in bulk MnTe is very small, which is the second key condition required to explain why the segregating In surfactant layers remains on the surface even for large layer thicknesses. For the MnTe\srf{} films, on the other hand, we only find Mn and Te at constant ratio throughout the films without traces of Sr or F atoms, except near the film/substrate interfaces, which confirms the higher purity of MnTe in this case.

\subsection{Electronic structure of the surface}

\begin{figure}[t]
	\centering
	\includegraphics[width=0.84\linewidth]{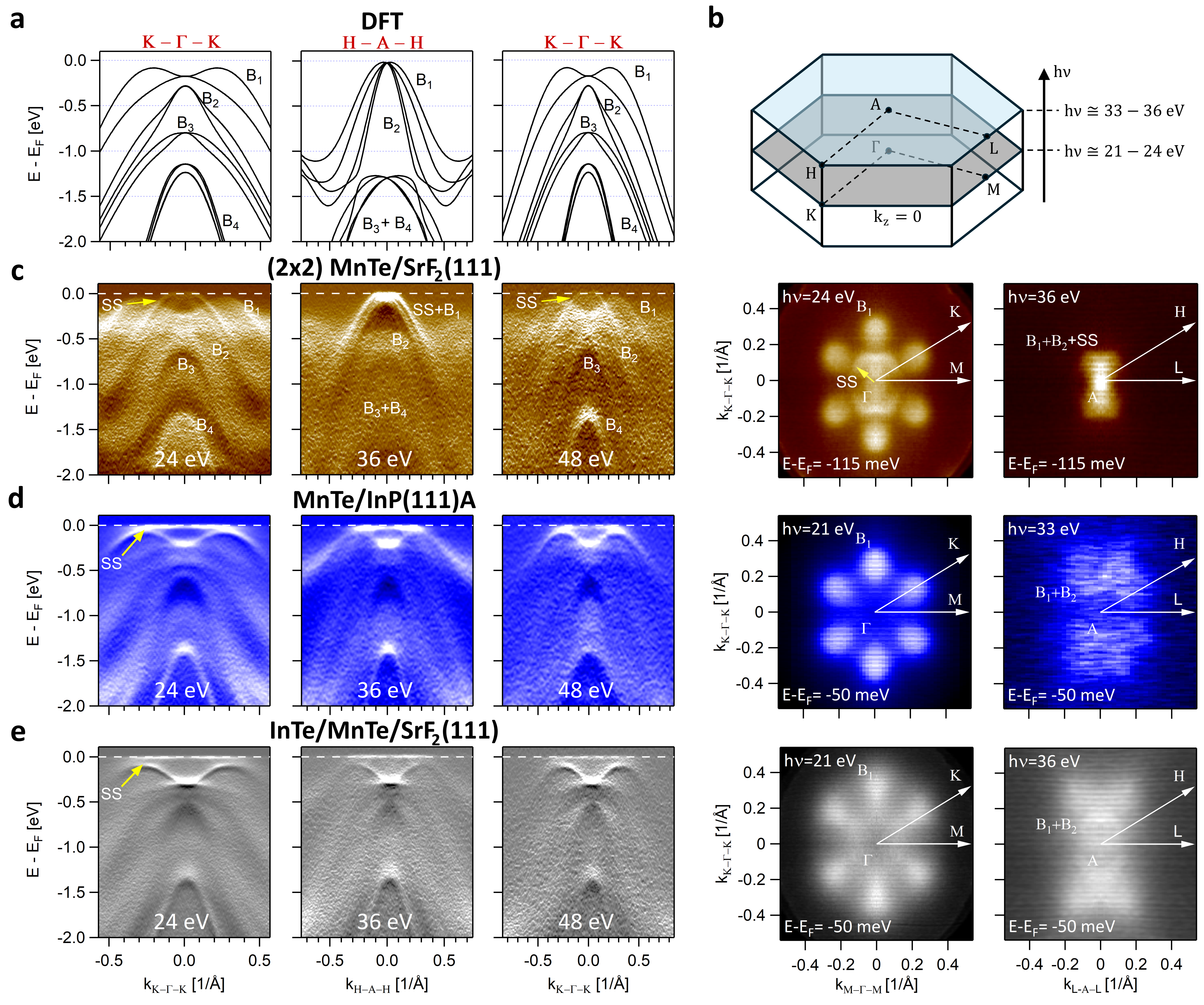}
	\caption{\textbf{ARPES measurements for MnTe (0001) with different surface terminations} compared to bulk DFT band structure calculations (a). 
    The $E(k)$ ARPES maps shown in (c-e) were recorded at 20 K for MnTe/$\text{SrF}_2$ (brown), MnTe/InP (blue) and InTe/MnTe/$\text{SrF}_2$ (grey)  using  photon energies of $h\nu$  = 24, 36, 48 that correspond to the $\Gamma$ and A points of the 3D Brillouin zone as shown in (b). The E(k) maps are displayed as first derivative and correspond to cuts along the K-$\Gamma$-K and H-A-H directions. On the right side, the corresponding constant energy maps (raw data) around the $\Gamma$ and A-points close to the Fermi level are depicted.  
    Note that the binding energies of the constant energy cuts are chosen to account for the slightly shifted Fermi levels between samples.}
	\label{fig:arpes}
\end{figure}

To determine the impact of the surface termination on the electronic structure, we performed ultraviolet angle-resolved photoemission spectroscopy (UV-ARPES) and spatially resolved scanning tunneling spectroscopy (STS). The ARPES results are presented in \textbf{Figure~\ref{fig:arpes}}c-e together with density functional theory (DFT) band structure calculations of bulk $\alpha$-MnTe shown in panel a for comparison. The $E(k)$ ARPES maps for MnTe/$\text{SrF}_2$ (brown data, panel c), MnTe/InP (blue data, panel d) and InTe/MnTe/$\text{SrF}_2$ (grey data, panel e)  were recorded at 30 K using photon energies of $h\nu$  = 24, 36 and 48 eV, corresponding to cuts along the K-$\Gamma$-K and H-A-H directions near the $\Gamma$ and A points \cite{Hajlaoui2024} of the 3D Brillouin zone as depicted in b. In addition, the corresponding constant energy maps (raw data) around $\Gamma$ and A near the Fermi level are shown on the right hand side. 

Evidently, the spectra of the clean (2$\times${}2) MnTe/\srf{} surface (panel c) differ significantly from those of the InTe-terminated MnTe surfaces (d and e) that in turn agree very well with each other.  
While we can identify the four characteristic bulk bands $B_1$-$B_4$ of altermagnetic MnTe by comparison with DFT calculations (panel a), in all three cases, significant differences emerge in the region near the Fermi level.
As indicated by the arrow, the clean (2$\times${}2) MnTe/\srf{} surface exhibits an additional surface-related  state (SS) near the Fermi level that is not present in the bulk calculations and whose spectral weight only weakly changes with photon energy.
This particular band is completely absent at the InTe-terminated MnTe/InP (panel d) and in InTe/MnTe/\srf{} (panel e) surfaces, which in contrast exhibit two other strong surface states that do not disperse noticeably with $k_z$ and are prominently present at both the $\Gamma$ and $A$ points.
The first one displays a pronounced camelback shape along K-$\Gamma$-K that mimics and overshadows the upper spin-split $B_1$ bulk band of MnTe. This is particularly evident in the 36 eV maps at the A-point, where the narrower MnTe $B_1$ and $B_2$ bulk bands are completely hidden behind this surface state.
The second feature sits directly atop the camelback state and crosses the Fermi level (as indicated by the arrows in d and e).
Due to their non-dispersive two-dimensional character, we conclude that they are connected to the InTe surfactant layer.

In the constant-energy Fermi surface cuts  (right panels of Figure~\ref{fig:arpes}c-e) near the $\Gamma$ point at 24 eV, the $B_1$ bulk valence band appears as six-fold symmetric lobes, consistent with previous investigations and indicative of averaging over multiple altermagnetic domains \cite{KrempaskyJ24N, Hajlaoui2024}. The additional surface band (SS) of the (2$\times${}2)-MnTe/\srf{} surface appears as a hexagonally shaped band inside these lobes, which persists also in the constant energy map at the A-point measured at 36 eV, where it overlaps with $B_1$ and $B_2$ bulk bands.
For the MnTe/InP (blue) and InTe/MnTe/\srf{} (grey) surfaces, this feature is absent in the constant energy maps, where instead, the two non-dispersive surface states overlapping with the bulk bands result in an elongation of the lobes at the $\Gamma$ point (24 eV) into oval shapes surrounded by a weak halo. Moreover, at the A-point at 36 eV, the constant energy maps reveal a wide hexagonal feature that is suppressed along the horizontal axis. The latter is mostly due to the strong InTe related surface state and contains only small contributions of the $B_1$ and $B_2$ bulk states. In addition, we note that the comparison of these maps suggests, that the Fermi level of the (2$\times${}2)-MnTe/\srf{} surface is slightly higher by about 65 meV, i.e., less p-type than for the InTe terminated surface. This is consistent with our finding that the (2$\times${}2) surface is less Te-rich, because only every second surface lattice site is occupied by Te anions. 

\begin{figure}[t]
	\centering
	\includegraphics[width=0.6\linewidth]{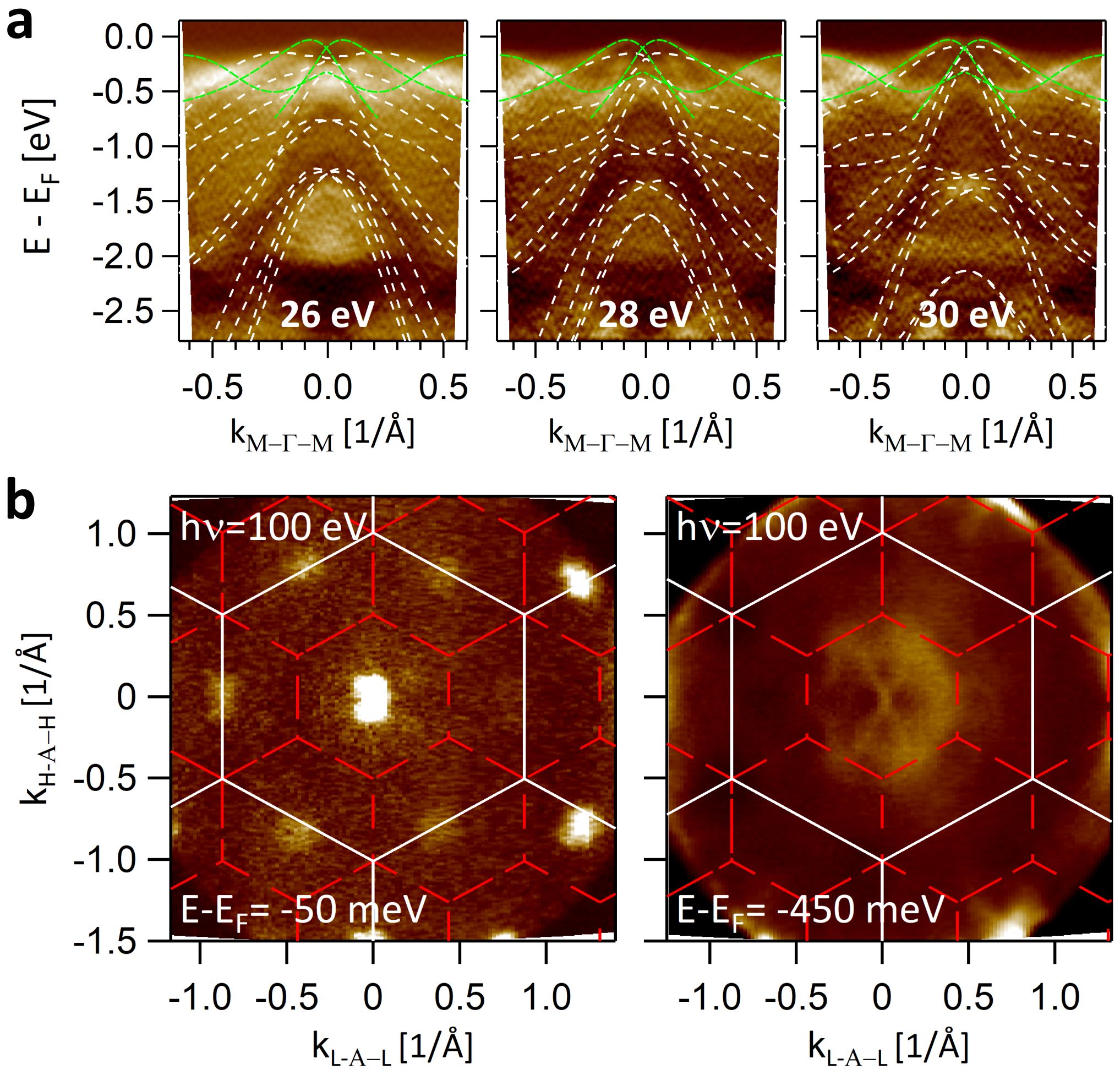}
	\caption{\textbf{Periodicity of the surface states of the (2$\times$ 2) MnTe/\srf{}(111) surface:}
    (a) Valence band spectra measured along the $M-\Gamma-M$ ($L-A-L$) direction at photon energies of $h\nu =$ 26 eV, 28 eV, 30 eV, corresponding to $k_z$ values between $\Gamma$ and $A$. Bulk band structure calculations (white dashed lines) and surface bands (green dashed lines as a ``guide to the eye'') close to the Fermi level are overlaid. (b) Constant-energy momentum distribution maps measured with $h\nu = 100$ eV, close to the $A$ point, at binding energies $E-E_F = -50$ meV (left) and $E-E_F = -450$ meV (right) with the (1$\times${}1) (white) and (2$\times${}2) surface BZ (red) overlaid onto the data. The surface bands near the Fermi level follow the (2$\times${}2) periodicity of the reconstructed surface (left), whereas the bulk features at higher binding energy follow the periodicity of the bulk lattice (right).}
	\label{fig:replica-bands}
\end{figure}

Due to the larger periodicity of the (2$\times${}2) reconstruction of MnTe/\srf{}(111), the surface Brillouin zone (BZ) of the surface band is expected to be reduced to \textit{half} of the projected bulk BZ. To check this point, we have recorded ARPES of this sample along the $M-\Gamma-M$ ($L-A-L$) direction between the $\Gamma$ and $A$ points ($h\nu $ between 24 and 36 eV) and constant energy maps at a higher photon energy of 100 eV to access a larger momentum space in $k_{x,y}$ directions. The results are shown in  \textbf{Figure~\ref{fig:replica-bands}}a and b, respectively, where in (a) the calculated bulk bands are overlaid as white dashed lines and the remaining  surface bands as green dashed lines as a ``guide to the eye''. Evidently, the surface band disperses strongly across the Brillouin zone with band maxima near 0 and $\pm 0.6\,\mathrm{\AA}^{-1}$.  
Moreover, the constant energy cut near the Fermi level ($E-E_F = -50$ meV, left hand side)
clearly exhibits replicas of the A-point with the periodicity given by reciprocal lattice of the $(2 \times 2)$ reconstruction. 

Conversely, the constant energy cut at deeper binding energies $E-E_F=-450$ meV shown on the right hand side of Figure~\ref{fig:replica-bands}b only exhibits the expected (1$\times${}1) bulk periodicity, evidencing that the lower energy bands in this momentum region are mostly bulk bands.

\begin{figure}[t]
	\centering
	\includegraphics[width=1.0\linewidth]{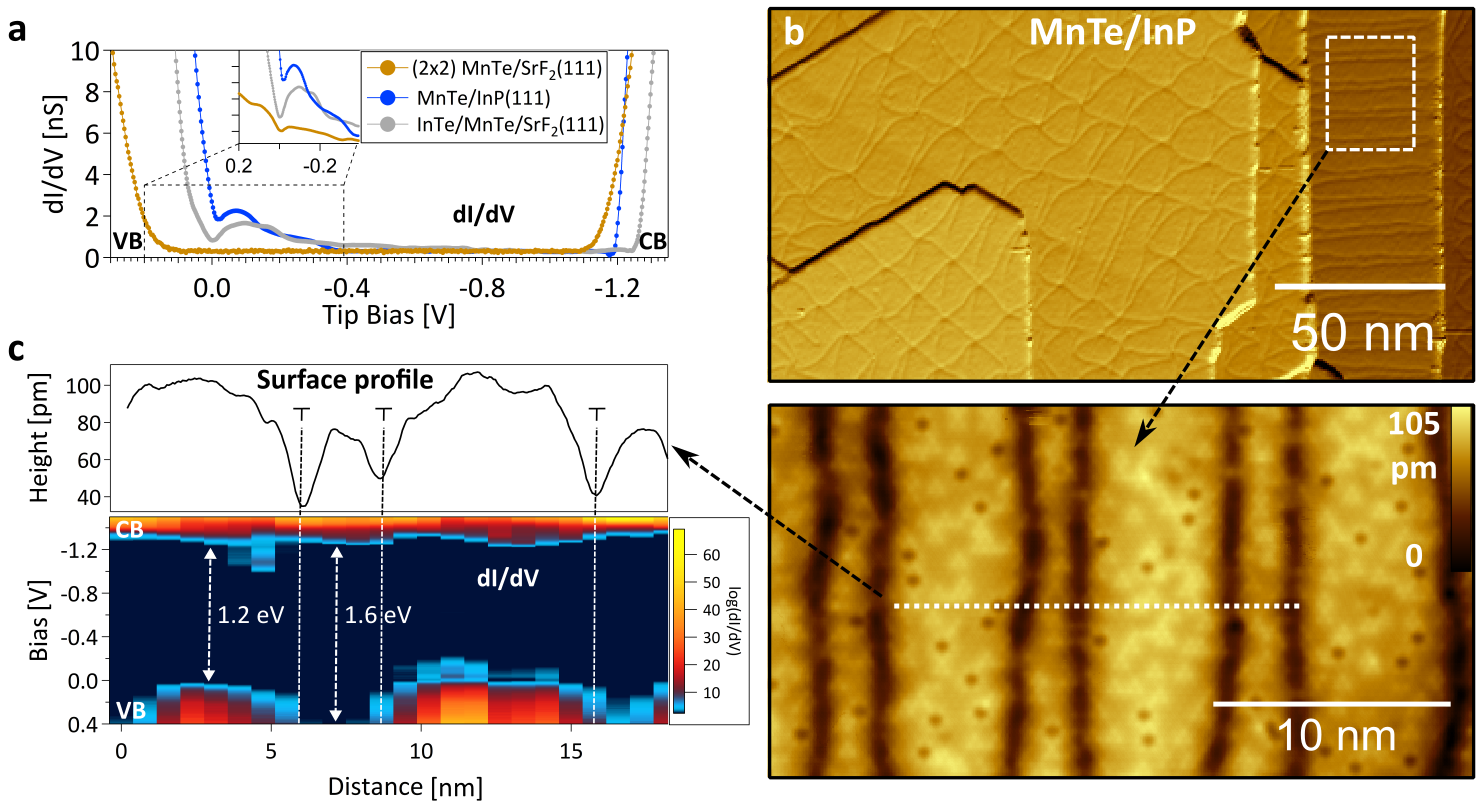}
	\caption{\textbf{Scanning tunneling spectroscopy of the MnTe surfaces at 4.4 K}. (a) STS spectra recorded in defect free areas on the MnTe/\srf{} (brown curve), MnTe/InP (blue curve) and MnTe/\srf{} with 1 ML InTe on top (grey curve) surfaces. Note that negative tip bias corresponds to tunneling into unoccupied states, meaning that at $V_{tip} < -1.2$ V tunneling occurs into the conduction band.  The zoomed-in inset reveals a weak non-zero dI/dV signal above zero bias that is almost negligible for MnTe/\srf{} but more pronounced for the InTe-terminated surfaces. (b) STM topography images of the MnTe/InP(111) surface ($V_{tip}=1.5$ V, $I_T = 500$ pA) with its characteristic dislocation network. The top image is shown in derivative mode to enhance the contrast of the dislocations. The dislocation pattern within the dashed rectangle is shown in the lower image on a magnified scale.  (c) STM surface profile (top) and color coded STS line map (bottom) measured  along the dotted line in (b), demonstrating the modulation of the electronic band gap due to the dislocation network. }
	\label{fig:sts-lines}
\end{figure}

For the InTe-terminated MnTe surfaces, we expect a local modulation of electronic properties due to the triangular dislocation network underneath the surface. Since this cannot be directly assessed by spatially averaging ARPES,  we have performed scanning tunneling spectroscopy (STS) at different locations of the surface at 4.4 K. \textbf{Figure~\ref{fig:sts-lines}}a compares the STS spectra of defect-free regions of the three different surface terminations.  For the clean (2$\times${}2) MnTe/\srf{} surface (brown curve), the local density of states (LDOS) mainly shows the onsets of the valence and conduction bands at about 0.15 and -1.1 eV, respectively. Note that negative tip bias means tunneling into unoccupied states. This yields a bulk band gap of $E_g \approx$ 1.25 eV, which agrees well with optical absorption measurements \cite{KriegnerD16NC,ALLEN1977367,PhysRevB.61.13679, Bossini_2020}.
In addition, we observe a faint non-zero LDOS above zero with a tail that extends to about -300 meV (see inset), which we associate with the surface band observed by ARPES.

The LDOS measured on the MnTe/InP (blue) and InTe/MnTe/\srf{} (grey) surfaces exhibits bulk valence and conduction band onsets close to 0 and -1.2 eV, respectively ({Figure~\ref{fig:sts-lines}}a), roughly amounting to the same bulk band gap. However, the onsets are slightly shifted to higher energies, which is consistent with the slightly lower Fermi level seen by ARPES for these samples. In addition, both spectra show clear LDOS peaks within the band gap above zero bias (see inset) at about 80 meV above the VB, extending to about 350 meV. This we attribute to the thin InTe surfactant layer that is expected to exhibit a smaller band gap than MnTe \cite{misra_InTe, Li_InTe-metallization}. 
We also note a small dip in the DOS at zero bias for all samples and that in the occupied states region at the VB , the DOS increases much more steeply for the InTe terminated surfaces. The position of the onset is compatible with the topmost bands observed by ARPES.

While the STS spectra of the (2$\times${}2) MnTe/\srf{} surface are uniform across the surface, the InTe-terminated MnTe surfaces display a pronounced variation of the band gap around the dislocations below the surface. This is demonstrated in Figure~\ref{fig:sts-lines}b and c by a spatially resolved STS map recorded in a line across several dislocations. 
In this map, the $dI/dV$ spectra are plotted in a color map as a function of distance along the line. The corresponding surface profile is shown on top to indicate the position of the dislocations. This reveals that around the dislocations the MnTe band gap is significantly increased to 1.6 eV compared to the 1.2 eV measured in between.
The modulation mainly occurs in the valence band (including surface states), whereas the conduction band onset remains relatively constant across the surface, only varying by few tens of meV. Most likely, this modulation is caused by the strong strain concentration at the dislocation cores that not only leads to the local surface deformations but also directly affects the band gap of the materials via the deformation potentials or the suppression of surface states.

\section{Conclusions}

By multiscale analysis  we have revealed the intricate atomic and electronic structure of altermagnetic $\alpha$-MnTe (0001) surfaces. Combining a wide range of surface sensitive techniques, we demonstrate that their properties strongly differ from idealized truncated bulk crystals. This is due to the natural tendency of the polar (0001) surface to reconstruct but also due to pronounced film/substrate interactions that for the MnTe growth on InP leads to an extraordinary In segregation and accumulation on the surface. The resulting InTe termination forms an ultra-stable, atomically thin surfactant layer that drastically lowers the free energy of the system. Most strikingly, at the InTe/MnTe interface a highly regular, quasi-periodic hexagonal dislocation network appears to accommodate the large InTe/MnTe lattice mismatch, which leads to strong strain variation in the layers. In contrast, the atomically clean MnTe surface obtained by growth on \srf{}(111) adopts a perfectly ordered (2$\times${}2) reconstruction in which only every second surface lattice site is occupied by a Te adatom. 

As shown by ARPES, while both surfaces display the spin-split altermagnetic electronic bulk bands, in addition, distinct surface states specific to each surface structure and surface termination appear. These are due, on the one hand, to the InTe surfactant layer but also emerge on the atomically clean MnTe(0001) surface where due to its (2$\times${}2) surface reconstruction the surface Brillouin zone is reduced to half of the surface-projected bulk Brillouin zone. This leads to the general conclusion that surface states add significant weight to the local density of states of altermagnetic MnTe, which due to their hybridization with the bulk bands are expected to significantly impact the altermagnetic spin textures and spin polarization probed by surface-sensitive probes such as ARPES and thus, needs to be taken into account in theory as well as experiments. Moreover, the increased density of states will directly affect the characteristics of thin film devices.
On the other hand, the dislocation network with its spatially varying band gap can provide a natural lateral confinement of the two-dimensional states that could be exploited for nanostructuring the surface or for tailoring the energy levels. 
Based on our results, similar effects are naturally expected for films grown on other substrate materials such as GaAs, Si or on van der Waals bonded V-VI chalcogenide buffer layers. 
While the wide variability of the atomic structure of MnTe surfaces and interfaces can be seen as a challenge, this also opens up a wide parameter space and engineering opportunity for tailoring the electronic and magnetic structure of altermagnets.

\section{Methods}

\subsection{Sample growth}
The samples were grown in a Varian GEN II molecular beam epitaxy system using elemental Mn and Te beam flux sources.  Epi-ready InP(111)A substrates were mounted on the MBE sample holders and the surface oxide removed by annealing at about 500 °C under ultra-high vacuum (UHV) conditions until a streaky pattern appeared in RHEED. Epitaxial  $\alpha$-MnTe films were then grown directly on the InP without any buffer layer. 
For \srf{}(111), the surface is quite stable in air so no oxygen desorption is required. 
For all samples, MnTe growth was performed at a substrate temperature around 360 °C, a Te/Mn beam flux ratio of about 3:1 and a growth rate of about 1 \AA/s, as determined by quartz microbalance beam flux measurements. Epitaxial growth was continuously monitored in situ by RHEED to ensure layer-by-layer growth, as depicted in Figure~\ref{fig:intro}b' \& c'. After the end of growth, the samples were rapidly cooled down to room temperature in order to preserve the epitaxial surface structure to avoid unintentional annealing.

For STM, LEED and XPS, the samples were transported without breaking UHV conditions in a Ferrovac vacuum suitcase to the PEARL beamline \cite{Pearl2017} of the Swiss Light Source. The same procedure was used for ARPES measurements at the URANOS beamline of the SOLARIS synchrotron in Krakow. The base pressure maintained in the vacuum suitcase was below $10^{-10}$ mbar, at which the surfaces were found to remain stable for several weeks. 

It is further noted, that for the samples on \srf{}, due to the highly insulating behavior of the substrate, a double growth procedure was used. In the first step, a MnTe buffer layer was grown and capped with tellurium. The samples were then removed from the vacuum, and tantalum stripes were attached over the sample edges to ensure good electrical contact to the thin film as needed for STM and photoemission measurements. In the second step, after reloading the sample to the MBE system, the Te cap was desorbed by annealing and another clean 200-500 nm thick MnTe film was grown on top to obtain a high surface quality.

\subsection{Photoelectron spectroscopy and LEED}

X-ray photoemission spectroscopy and LEED were performed at the PEARL beamline \cite{Pearl2017} of the Swiss Light Source. 
LEED was measured with an Omicron SpectaLEED unit. Electron beam energies ranging from 20 to 100 eV were used to characterize the sample surfaces.
XPS spectra were recorded using the $\text{K}_\alpha$ emission line of a Mg anode (photon energy 1253.6 eV) and a Scienta DFS30 analyser in transmission mode.
For quantifying elemental ratios, the Mn 2p, Te 3p and In 3p core-level peaks were measured at $\theta = 20\degree$ and $\theta = 60\degree$ relative to the surface normal. 
The angle between incoming x-rays and electron detection was fixed at $60\degree$.
The weight of the respective peaks were extracted from curve fits using Gaussian profiles for the peaks and cubic polynomials for the background subtraction. 
The spectra and details about the data analysis are presented in the supplementary material.

\subsection{Scanning tunneling microscopy and spectroscopy}

Scanning tunneling microscopy and spectroscopy was performed using PtIr tips in an Omicron low-temperature STM operated at 4.4 K under ultrahigh vacuum conditions at the PEARL beamline \cite{Pearl2017}.
Bias voltages were applied to the tip, such that applied positive bias probes the occupied electronic states and negative bias the unoccupied states of the sample.
The vacuum chambers of the STM, XPS and LEED stations were all connected with UHV transfers.

\subsection{Angle-resolved photoemission spectroscopy}

ARPES was measured at the Ultra-Resolved ANgular phOtoelectron Spectroscopy (URANOS) beamline of the SOLARIS synchrotron \cite{uranos} in Poland. ARPES spectra were recorded using p-polarized light in the photon energy range of 21 eV to 100 eV at a temperature below 30 K -- well below the magnetic phase transition temperature of 306 K. Photoelectrons were analyzed  with a VG Scienta DA30L electron spectrometer. Angular and energy resolution are estimated to be about 0.1\degree\ and 7.5 meV, respectively.

\subsection{Secondary ion mass spectrometry}

SIMS measurements were performed ex situ with an IONTOF SIMS5 time-of-flight spectrometer at Brno University of Technology, Czech Republic. Analysis was performed using Bi ions with an energy of 30 keV in an area of $100 \times 100$ $\mu m^2$. For depth profiling, the films were sputtered by oxygen ions with an energy of 2 keV in a larger area of $220 \times 220$ $\mu m^2$. Positive ions were detected to determine the elemental concentrations of the cations (Mn, In and Sr) in the sample. In addition, the signal of the anion elements (F, Te and P) was also detected.

\subsection{Band structure calculations}

The self-consistent electronic structure calculation of the magnetically ordered phase of bulk $\alpha$-MnTe with lattice parameters $a=4.15$ Å and $c=6.71$ Å was performed using the Vienna Ab initio Simulation Package (VASP) \cite{KRESSE199615} utilizing projector-augmented-wave pseudopotentials \cite{bloechl_projector} as described in our previous work \cite{Hajlaoui2024}. The exchange-correlation functional GGA was chosen according to \cite{perdew_burke}. Brillouin zone sampling was performed on a $\Gamma$-centered $9 \times 9 \times 4$ mesh, and a consistent energy cutoff of 400 eV was applied in all calculations. According to previous estimates from \cite{mu_phonons-mnte}, the screened on-site Coulomb interaction $U$ and exchange interaction $J$ for Mn were set to 4.80 and 0.80 eV respectively. Spin orbit coupling was included on top of the Schrödinger Hamiltonian and the Néel vector direction was set along the $[1\overline{1}00]$ direction, in accordance with experimental evidence from neutron diffraction and magneto-transport measurements \cite{KriegnerD17PRB}.  

\medskip
\textbf{Supporting Information} \par 
Supporting Information is available from the corresponding authors.

\medskip
\textbf{Acknowledgements} \par 
The authors thank Juraj Krempask\'{y} and Thomas Jung for helpful discussions. M.H. acknowledges funding by the Swiss Nanoscience Institute. G.S. and M.H. acknowledge support by the Austrian Science Fund (FWF, Projects PIN 6540324 and I06568) and the University of Linz (SEED-Project LIT-2022-11-131). 
We acknowledge the SOLARIS synchrotron in Krakow, Poland, for the provision of synchrotron radiation beamtime at the URANOS beamline. Research at the National Synchrotron Radiation Centre SOLARIS is supported by the Ministry of Science and Higher Education, Poland, under contract no. MNiSW/2026/DWM/928.

\newpage

%
\bibliographystyle{MSP}
\bibliography{mnte}

\end{document}